\documentclass[aip,rsi,reprint]{revtex4-1}
\usepackage{siunitx}
\usepackage{graphicx}
\usepackage{hyperref}
\usepackage{amssymb}

\hypersetup{ 
           colorlinks=true,
           citecolor=blue, 
           filecolor=blue,   
            linkcolor=blue,    
            anchorcolor=blue,
            urlcolor=blue,  
            pdfborder=0 0 1,   
}

\begin{document}


\title{A cavity optomechanical locking scheme based on the optical spring effect} 



\author{P. Rohse}
\affiliation{ZOQ (Zentrum f\"{u}r Optische Quantentechnologien)\\Universit\"{a}t Hamburg, Luruper Chaussee 149, 22761 Hamburg, Germany}
\author{J. Butlewski}%
\affiliation{ZOQ (Zentrum f\"{u}r Optische Quantentechnologien)\\Universit\"{a}t Hamburg, Luruper Chaussee 149, 22761 Hamburg, Germany}
\author{F. Klein}%
\affiliation{ZOQ (Zentrum f\"{u}r Optische Quantentechnologien)\\Universit\"{a}t Hamburg, Luruper Chaussee 149, 22761 Hamburg, Germany}
\author{T. Wagner}%
\affiliation{ZOQ (Zentrum f\"{u}r Optische Quantentechnologien)\\Universit\"{a}t Hamburg, Luruper Chaussee 149, 22761 Hamburg, Germany}
\author{C. Friesen}
\affiliation{INF (Institut f\"{u}r Nanostruktur- und Festk\"{o}rperphysik)\\Universit\"{a}t Hamburg, Jungiusstra{\ss}e 9, 20355 Hamburg, Germany}
\author{A. Schwarz}
\affiliation{INF (Institut f\"{u}r Nanostruktur- und Festk\"{o}rperphysik)\\Universit\"{a}t Hamburg, Jungiusstra{\ss}e 9, 20355 Hamburg, Germany}
\author{R. Wiesendanger}
\affiliation{INF (Institut f\"{u}r Nanostruktur- und Festk\"{o}rperphysik)\\Universit\"{a}t Hamburg, Jungiusstra{\ss}e 9, 20355 Hamburg, Germany}
\affiliation{ZOQ (Zentrum f\"{u}r Optische Quantentechnologien)\\Universit\"{a}t Hamburg, Luruper Chaussee 149, 22761 Hamburg, Germany}
\author{K. Sengstock}
\affiliation{ZOQ (Zentrum f\"{u}r Optische Quantentechnologien)\\Universit\"{a}t Hamburg, Luruper Chaussee 149, 22761 Hamburg, Germany}
\affiliation{ILP (Institut f\"{u}r Laserphysik)\\Universit\"{a}t Hamburg, Luruper Chaussee 149, 22761 Hamburg, Germany}
\author{C. Becker}
\email{cbecker@physnet.uni-hamburg.de}
\affiliation{ZOQ (Zentrum f\"{u}r Optische Quantentechnologien)\\Universit\"{a}t Hamburg, Luruper Chaussee 149, 22761 Hamburg, Germany}
\affiliation{ILP (Institut f\"{u}r Laserphysik)\\Universit\"{a}t Hamburg, Luruper Chaussee 149, 22761 Hamburg, Germany}



\date{\today}

\begin{abstract}
We present a novel locking scheme for active length-stabilization and frequency detuning of a cavity optomechanical device based on the optical spring effect.
The scheme can be used as an alternative to the Pound-Drever-Hall locking technique but in contrast doesn't require signal processing on time-scales of the cavity decay rate.
It is therefore particularly  suited for stabilizing micro cavities, where this time-scale can be extremely fast. 
The error signal is generated through the optical spring effect, i.e. the detuning-dependent frequency-shift of a nanomechanical oscillator that is dispersively coupled to the intra-cavity light field.
We explain the functional principle of the lock and characterize its performance in terms of bandwidth and gain profile. 
The optical spring locking scheme can be implemented without larger efforts in a wide variety of optomechanical systems in the unresolved sideband regime. 
\end{abstract}


\maketitle 

\section{Introduction}
\label{sec:Introduction}
In the last decades, the field of cavity optomechanics \cite{Aspelmeyer2014, WarwickP.Bowen2015} has made impressive advances in combining mechanical systems with the powerful tools of quantum optics. This rapid development has lead to many new ideas for precision sensing\cite{Abbott2016,Moller2017,Reinhardt2016}, a
wide variety of quantum technological applications \cite{Kurizki2015,Rogers2014} and fundamental tests of quantum mechanics in a completely new parameter regime
of macroscopic objects\cite{Purdy2013,Riedinger2018,Ockeloen-Korppi2018}.

In the prototypical optomechanical system a mechanical mode interacts with an optical cavity mode driven by a laser field. In order to pump the cavity in a controlled way, the detuning $\Delta = \omega_\mathrm{L} - \omega_\mathrm{c}$ between laser field and cavity typically has to be set to a definite value close to the cavity resonance. The exact value of the detuning determines the dynamical back-action of the optical mode onto the mechanical oscillator, given by the phase-delay between mechanical motion and intra-cavity field\cite{Aspelmeyer2014}. While the out-of-phase quadrature of the field leads to a radiation-pressure mediated energy transfer in terms of heating or cooling, the in-phase quadrature effectively alters the mechanical spring constant and shifts the frequency of the mechanical mode, known as the optical spring effect. For example, a resonant laser drive $\Delta = 0$ can be used for precision sensing of the mechanical motion\cite{Purdy2013}, while a red detuned drive $\Delta < 0$ is routinely used to cool the mechanical mode into the quantum realm. This method known as sideband-cooling has become a standard technique for cooling a mechanical mode into its quantum ground-state\cite{Chan2011,Teufel2011,Peterson2016}. A blue detuned drive $\Delta > 0$ heats the mechanical mode and is usually avoided in optomechanical experiments. 

\begin{figure} 
  \centering
    \includegraphics{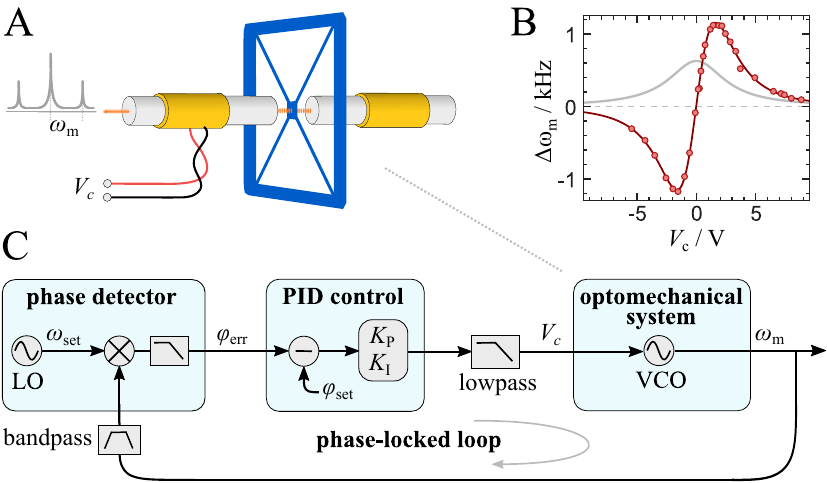}
         \caption{\label{fig:optolock} \textbf{Scheme of the optical spring-control.} 
         Our fiber-based optomechanical MiM cavity\cite{Zhong2017} (\textbf{A}) is length-controlled to lock the detuning $\Delta$ relative to the laser drive. 
         For this, the control voltage $V_\mathrm{c}$ is applied to one of the two piezos (yellow tubes), which allow for scanning both cavity mirrors independently. 
         The motion of the SiN trampoline\cite{Norcada} oscillator (blue) imprinting sidebands of frequency $\omega_\mathrm{m}$ on the laser is measured through balanced homodyne detection (not shown). 
         The locking error signal is derived from the optical spring effect (\textbf{B}), a detuning-dependent frequency-shift $\Delta \omega_\mathrm{m} \left( V_\mathrm{c} \right)$, which is shown for a cavity input power of $P_\mathrm{in} = 43 \,$\si{\micro}W. 
         This voltage-controlled oscillator (VCO) is used in a digital phase-locked-loop (\textbf{C}) to lock the trampoline frequency $\omega_\mathrm{m}$ to a set frequency $\omega_\mathrm{set}$. }
\end{figure}

However, pulsed experiments with blue detuned laser pulses have demonstrated entanglement and Fock-state generation of the mechanical mode\cite{Palomaki2013,Riedinger2018}. Furthermore, the coherent amplification of mechanical quanta for a blue detuned drive is the underlying process in a phonon laser\cite{Vahala2009}.

In our hybrid experiments\cite{Christoph2018} we operate an optomechanical system close to resonance, which we achieve using a new optomechanical locking scheme based on the optical spring effect (see Figure \ref{fig:optolock}). The relevance of this technique will be motivated in the following.

In order to lock the detuning between the driving laser and the cavity mode, either the mode frequency $\omega_\mathrm{c}$ of the cavity needs to be actively stabilized, or the laser frequency $\omega_\mathrm{L}$ needs to be stabilized with respect to the cavity. In both cases, a suitable locking error signal needs to be derived. The simplest solution is the so-called side-of-fringe lock, where the error signal is given by the cavity response itself -- the detuning-dependent transmission or reflection signal. For example, if the cavity transmission is stabilized to 50\% of the on-resonance maximum, the detuning is locked to the value of the cavity's HWHM linewidth $\kappa = \gamma/2$. One major drawback of this scheme is that the locking range is restricted to non-zero detunings, where the cavity response is not at its maximum and has a finite slope. Therefore, side-of-fringe locking is not suitable in our hybrid experiment, as we aim at maximizing the coupling of the optomechanical system to an ultra-cold atomic cloud\cite{Christoph2018}.

In contrast to the side-of-fringe lock, the powerful and widely used Pound–Drever–Hall technique (PDH)\cite{Drever1983,Black2001} allows for locking at zero detuning with a much steeper error signal leading to a significantly more robust lock. This is because the error signal is connected to the cavity's phase response, which exhibits a slope on resonance. However, due to several reasons PDH locking is difficult in our specific setup\cite{Zhong2017}. Firstly, the spectral linewidth of our cavity is extremely large and on the order of $\gamma = 2 \pi \, \times  48 \,  \mathrm{GHz}$ (see Section \ref{sec:FunctionalPrinciple}). As the PDH technique requires optical sideband generation and signal processing at frequencies on the order of $ \gamma$, this is a costly and technically demanding task in our system. Secondly, our fiber cavity is asymmetric\cite{Bick2016} (or single-sided) and we would have to couple the PDH beam from the highly-reflecting side into the cavity. Since the cavity's phase profile on this side is much less pronounced than in the normal case for a symmetric cavity, the resulting PDH signal is very weak. As a result, we decided to generate our locking error signal in a different way.

Light-matter interactions inside a cavity offer a variety of opportunities for locking techniques. For example, self-locking and also active control through thermal expansion have been described, which is most relevant in micro-cavities\cite{Carmon2005, Gallego2016}. Furthermore, radiation-pressure mediated self-locking is routinely used in gravitational wave interferometers. Specifically, a strong optical spring effect is used to increase the mechanical resonance frequency to larger values above seismic perturbations\cite{Corbitt2006,LIGOScientificCollaboration2009}. In addition to that, radiation-pressure can be used as an actuator in a feedback loop, which can significantly increase the locking bandwidth compared to the usual piezo-actuation\cite{McClelland2018}.

Our scheme marks a new approach by using the optical spring effect as the error signal for an active lock. It can be implemented with minimal technical efforts as most of the required components are typical elements of an optomechanical system.

\section{Functional principle}
\label{sec:FunctionalPrinciple}
This Section explains the functional principle of the optical spring-control. The active feedback loop uses the cavity detuning-dependent frequency-shift by the optical spring as the error signal and a cavity mirror piezo as the feedback actuator to control the cavity length. The loop diagram of this optomechanical locking scheme is shown in Figure \ref{fig:optolock}C. As a piezo actuation of voltage $V_\mathrm{c}$ leads to a detuning-related frequency-shift $\Delta \omega_\mathrm{m} \left( V_\mathrm{c} \right)$ through the optical spring (see Figure \ref{fig:optolock}B), the whole cavity optomechanical system can be regarded as a voltage-controlled oscillator (VCO). The control loop uses this VCO in a phase-locked loop (PLL) to lock the frequency $\omega_\mathrm{m}$ of the mechanical oscillator -- and hence the detuning of the cavity -- to a digital local oscillator (LO) of frequency $\omega_\mathrm{set}$. In the following, the individual parts of this control loop will be described.

\textit{\textbf{The optomechanical system}}: Our cavity optomechanical device consists of a cryogenic all-fiber cavity inside a dilution refridgerator providing a continuous bath temperature of $500 \, \mathrm{mK}$\cite{Zhong2017}. The fiber cavity is asymmetric\cite{Bick2016} (or single-sided) to provide a finite reflectivity on resonance, which is required for a bi-directional interaction in our hybrid atomic-mechanical experiments\cite{Christoph2018}. As a result, the low-reflective side of the cavity limits the cavity finesse to small values -- currently we work with $\mathcal{F}=134$ (at the point of largest optomechanical coupling \cite{Wilson2009}). Together with the likewise small cavity length $L = 23.3 \,$\si{\micro}m the resulting spectral linewidth is extremely large, namely $ \gamma = \pi c/( L \mathcal{F}) =  2 \pi \, \times  48 \,  \mathrm{GHz}$. As mentioned in the previous section, this makes PDH locking a challenging task in our system. The mechanical oscillator is a state-of-the-art trampoline oscillator\cite{Reinhardt2016,Norte2016, Norcada} with a fundamental mode at $\omega_\mathrm{m} = 2 \pi \times 154 \, \mathrm{kHz}$ with an effective mass of $3 \, \mathrm{ng}$ and an ultra-high Q-factor of $8.9 \times 10^7$. It is placed inside the fiber cavity, forming a so-called membrane-in-the-middle\cite{Jayich2008,Wilson2009} (MiM) configuration (see Figure \ref{fig:optolock}A). Through approaching the trampoline very close ($4.5 \,$\si{\micro}m) to one of the cavity fibers we reach an optomechanical coupling rate $g_\mathrm{m} =  2 \pi \, \times  16.4 \, \mathrm{GHz/nm}$ with a single-photon coupling rate of $g_\mathrm{0} = 2 \pi \, \times  69.7 \, \mathrm{kHz}$.

Since $ \kappa \gg \omega_\mathrm{m}$, the optomechanical system operates far in the unresolved sideband regime and the frequency-shift $\Delta \omega_\mathrm{m}$ by the optical spring effect takes the simple form shown in Figure \ref{fig:optolock}B. The measured data was fitted using the expression for the optical spring effect\cite{Aspelmeyer2014}
\begin{equation}
\label{eq:Delta_omega}
\Delta \omega_\mathrm{m} = \bar{n} g_\mathrm{0}^2 \left[ \frac{\Delta - \omega_\mathrm{m}}{\kappa^2 + \left( \Delta - \omega_\mathrm{m} \right)^2} + \frac{\Delta + \omega_\mathrm{m}}{\kappa^2 + \left( \Delta + \omega_\mathrm{m} \right)^2} \right] 
\end{equation}
where $\bar{n} = \bar{n}_\mathrm{0} /(1+(\Delta/\kappa)^2)$ is the  detuning-dependent intra-cavity photon number. Note that this frequency-shift has a different functional behavior if the system enters the resolved sideband regime and even vanishes around the cavity resonance for $ \kappa \ll \omega_\mathrm{m}$.  

Another important consequence of the extremely fast cavity decay rate $\gamma$ is that the intra-cavity field (and correspondingly the radiation-pressure induced optical spring) follows the detuning adiabatically on all time scales relevant for our locking scheme. Hence, the frequency-dependence of the optomechanical VCO can be neglected in the loop analysis. Also the gain and phase of the cavity piezos (which were measured to be flat below $10 \, \mathrm{kHz}$) and the balanced homodyne detector (bandwidth $1 \, \mathrm{MHz}$, part of the VCO element in Figure \ref{fig:optolock}C) can be neglected.

\textit{\textbf{The phase detector}}: We typically use a digital lock-in amplifier\cite{ZI} to analyze the motional state of our mechanical oscillator. Its amplitude and phase are obtained through dual-phase demodulation of the signal from our balanced homodyne detector (part of the VCO element in Figure \ref{fig:optolock}C). Such a demodulator also serves as the phase detector for the lock. Specifically, the phase of the VCO is measured with respect to the LO of the demodulator. Hence, the demodulator LO frequency $\omega_\mathrm{set}$ determines the set frequency of the PLL (and the mechanical oscillator, respectively). Since the demodulator bandwidth $B_\mathrm{d}$ limits the gain and phase response of the detector, it was set to $B_\mathrm{d} = 10 \, \mathrm{kHz}$ which is much larger than the locking bandwidth (as shown later). The demodulator phase output $\varphi_\mathrm{err}$ is the error signal for the PID controller. It is output in the form of an analog voltage, scaled by $10 \, \mathrm{mV/deg}$. The analog input of the detector is equipped with a passive $150 \, \mathrm{kHz}$ bandpass filter (passband bandwidth $50 \, \mathrm{kHz}$).

\textit{\textbf{The PID controller}}: The PID control is realized with a built-in function of the digital lock-in amplifier\cite{ZI}. It is fed with the error signal $\varphi_\mathrm{err}$ from the phase detector and compares it with the set phase $\varphi_\mathrm{set} = 0$. The voltage P-gain is set to $K_\mathrm{P} = 3$ and the voltage I-gain is set to $K_\mathrm{I} = 50 \, \mathrm{Hz}$. The integrator part is used to shape the loop phase response below $10 \, \mathrm{Hz}$ to optimize the steady state stability. The controller output is an analog voltage which is fed to one of the MiM cavity piezos. It passes a high-order lowpass filter (bandwidth $5 \, \mathrm{kHz}$) which prevents noise from disturbing the mechanical oscillator. The filter is not an essential part of the lock and can be omitted.

\section{Performance and characterization}
\label{sec:Performance and characterization}

In this section, the performance of the spring-control is analyzed in terms of locking bandwidth and stability. For this, the response of the control to an external perturbation was investigated. We show that this closed-loop response is precisely predictable through measuring the transfer functions of all relevant loop elements. Finally, we discuss the practicality and feasibility of our active spring-control in other optomechanical systems.

For practical reasons, the characterization measurements in this section were performed by locking the spring-control PLL to the fundamental trampoline s1-mode at $\omega_\mathrm{m} = 2 \pi \times 154\, \mathrm{kHz}$\cite{Reinhardt2016}. However, if this mode is used in an experiment, the lock can be operated with a higher trampoline mode.

\textit{\textbf{Loop analysis}}: Typically, the performance of a control loop is evaluated by measuring the lock's capability of compensating for an external perturbation. This so-called closed-loop response $G_\mathrm{cl}$ is determined by the gains and phases of the elements in the control loop which can be summarized in one quantity, the so-called open-loop response $G_\mathrm{0}$. This complex-valued transfer function is the product of all transfer functions in the loop:
\begin{equation}
G_\mathrm{0} \,  = \, G_\mathrm{det} \, \cdot \, G_\mathrm{PID} \, \cdot \, G_\mathrm{VCO} \, \cdot \, G_\mathrm{lp} \, \cdot \, G_\mathrm{bp} 
\label{eq:G_0}
\end{equation}
with the individual open-loop transfer functions $G_\mathrm{det}$\cite{G_det} from the phase detector, $G_\mathrm{PID}$ from the PID controller, $G_\mathrm{VCO}$ from the optomechanical MiM cavity and $G_\mathrm{lp}$ ($G_\mathrm{bp}$) from the passive lowpass (bandpass) filter (the frequency-dependence was omitted for clarity). Each individual transfer function $G_\mathrm{i}(s)$ describes the frequency-dependent gain $|G_\mathrm{i}(s)|$ and phase delay $\mathrm{arg}[G_\mathrm{i}(s)]$ of the corresponding loop element.

\begin{figure} 
  \centering
    \includegraphics{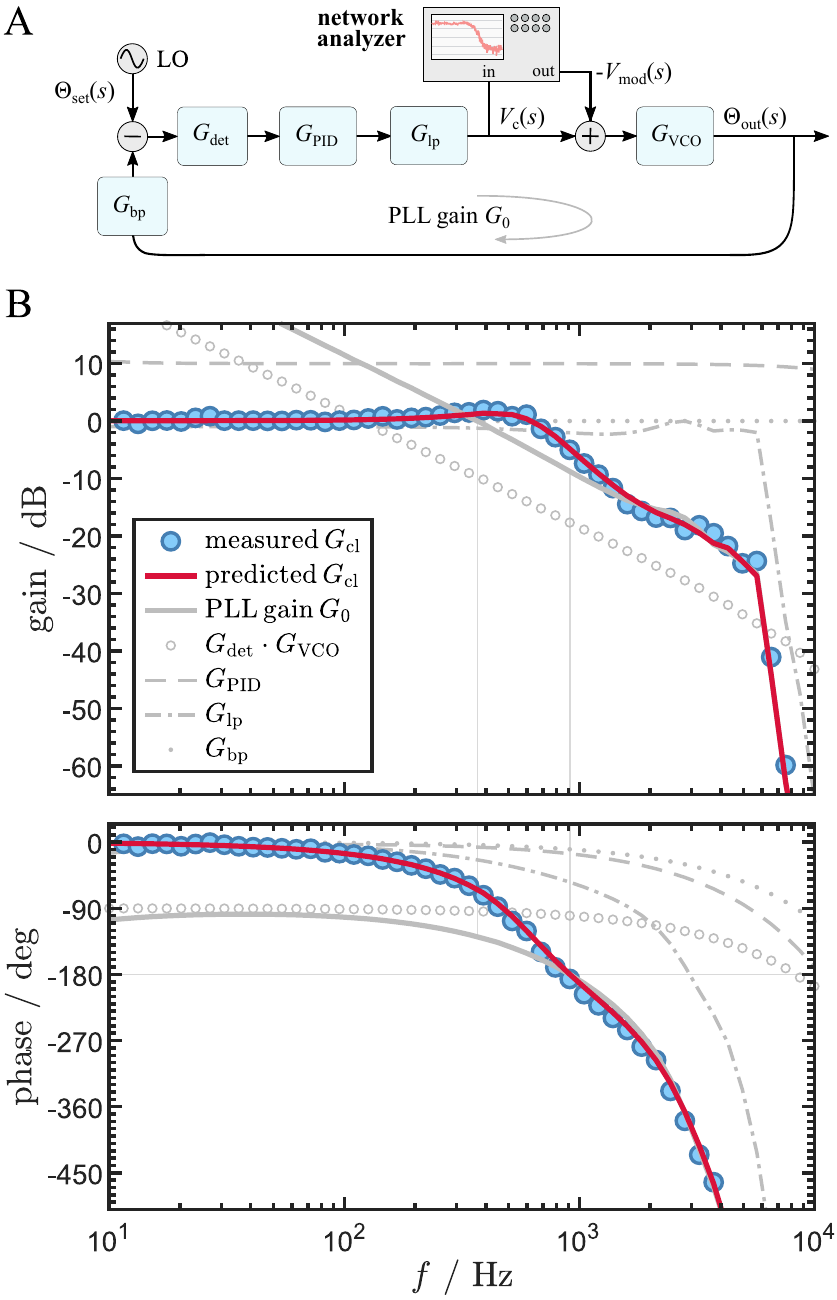}
     \caption{\label{fig:loopresponse} \textbf{Loop response analysis.} Direct measurement of closed-loop response $G_\mathrm{cl}$ and indirect measurement of open-loop response $G_\mathrm{0}$ through measurement of responses $G_\mathrm{i}$ of the individual PLL elements. (\textbf{A}): Linear system block diagram of the PLL in the Laplace domain\cite{PLL} with the measurement arrangement to acquire $G_\mathrm{cl}$. (\textbf{B}): Gain $|G|$ and phase $\mathrm{arg}(G)$ of measured $G_\mathrm{cl}$ (blue circles) and the predicted curve (red solid line) calculated with expression \ref{eq:G_cl} using the measured $G_\mathrm{0}$ (gray solid line). $G_\mathrm{0}$ was obtained by measuring the individual response functions $G_\mathrm{det}$ of the phase detector, $G_\mathrm{PID}$ of the PID control and $G_\mathrm{lp}$ ($G_\mathrm{bp}$) from the passive lowpass (bandpass) filter. $G_\mathrm{VCO} \in \mathbb{R}$ is the only free parameter in $G_\mathrm{0}$ to fit $G_\mathrm{cl}$. The unity gain frequency $f_\mathrm{0} = 370 \, \mathrm{Hz}$ (logarithmic zero-crossing at $|G_\mathrm{0}| =1$) and $f_\mathrm{\pi} = 910 \, \mathrm{Hz}$ for positive feedback ($\mathrm{arg}(G_\mathrm{0}) = -180^{\circ}$) are indicated by vertical lines.}
\end{figure}

To acquire the closed-loop response of our spring-control PLL, we generate a phase modulation by applying a modulation voltage $ V_\mathrm{mod}$ at one cavity piezo and measure the controller output voltage $ V_\mathrm{c}$. The corresponding closed-loop response function $G_\mathrm{cl} = V_\mathrm{c} / V_\mathrm{mod}$ can be read off from the block diagram in the Laplace domain\cite{PLL} (see Figure \ref{fig:loopresponse}A): $V_\mathrm{c}(s) = -  G_\mathrm{0}(s) [ V_\mathrm{c}(s) - V_\mathrm{mod}(s)]$. This yields:
\begin{equation}
G_\mathrm{cl}(s) \,  = \, \frac{V_\mathrm{c}(s)}{V_\mathrm{mod}(s)} \, =  \,  \frac{G_\mathrm{0}(s)}{1 + G_\mathrm{0}(s)} \, .
\label{eq:G_cl}
\end{equation}
Here, the LO set phase $\Theta_\mathrm{set}$ was set to zero, yielding the well-known expression for the closed-loop response $G_\mathrm{cl}$ of a negative feedback loop. 

The measurement arrangement for acquiring the closed-loop response $G_\mathrm{cl}$ is shown in Figure \ref{fig:loopresponse}A. The loop analysis was performed with the built-in network analyzer (NA) function of the lock-in amplifier\cite{ZI}. At first, the MiM cavity was stabilized at a red detuned laser drive $\Delta < 0$, by locking the trampoline frequency to $\omega_\mathrm{m} = 2 \pi \times 154.2 \, \mathrm{kHz}$ with the spring-control PLL. While the lock was active, the stabilized cavity length was disturbed by connecting the NA output voltage $-V_\mathrm{mod}$ to a cavity piezo. Then, the output control voltage $V_\mathrm{c}$ of the control loop was picked off for the NA input signal. Technically, the clock-wise direction of the signal flow in Figure \ref{fig:loopresponse}A was fulfilled by connecting the NA output to the second cavity piezo that is not part of the control loop (see Figure \ref{fig:optolock}A). This is physically equivalent to Figure \ref{fig:loopresponse}A, except for a constant prefactor of $G_\mathrm{cl}$ (depending on asymmetries in the MiM cavity like mirror reflectivities or piezo gains), which was removed by normalizing the measured $G_\mathrm{cl}$ to the first data points around $f = 10 \, \mathrm{Hz}$. The minus-sign of $-V_\mathrm{mod}$ is just a result of the polarity of the cavity piezo in this specific arrangement. The measured closed-loop gain $|G_\mathrm{cl}|$ and phase $\mathrm{arg}(G_\mathrm{cl})$ are shown in Figur \ref{fig:loopresponse}B.

According to equation \ref{eq:G_cl}, the measured $G_\mathrm{cl}$ can be fully described by the open-loop response $G_\mathrm{0}$. The problem of measuring $G_\mathrm{0}$ is that it is practically impossible to measure the open-loop response of a PLL. This is because any NA measurement would require extreme stability of the free running VCO and detector LO (which is generally not the case) since otherwise the loop phase and correspondingly the gain would be completely undefined. However, technical workarounds have been reported in the literature. For example, the bandwidth of the PLL can be strongly reduced to measure a quasi open-loop response out of the locking bandwidth\cite{Ye2017}.

Here, we follow a different approach and exploit that our phase detector is a digital LO with an extremely good frequency stability. Hence, the open-loop transfer function $G_\mathrm{det}$ of the phase detector can be measured by using another digital test VCO with equal frequency stability. Technically, $G_\mathrm{det}$ (and $G_\mathrm{bp}$) were measured by frequency-modulation of such a test VCO with the NA output and feeding the frequency-modulated VCO signal into the phase detector. Then, the phase detector output was fed into the NA input in form of an analog voltage. The detected phase is the integral of the frequency change and thus the phase detector behaves as an integrator: $|G_\mathrm{det}|$ has the typical slope of -20 dB per decade and $\mathrm{arg}(G_\mathrm{det}) = -90^{\circ}$\cite{G_det_2}. The additional phase delay $\mathrm{arg}(G_\mathrm{det}) < -90^{\circ}$ for $f > 10 \, \mathrm{kHz}$ originates from the bandwidth setting $B_\mathrm{d} = 10 \, \mathrm{kHz}$ of the demodulator (a larger demodulator bandwidth generates more phase noise). Notably, the parasitic phase delay of the test VCO was subtracted out by increasing the bandwidth of the detection demodulator to $B_\mathrm{d} = 50 \, \mathrm{kHz}$ where the frequency-dependence of $\mathrm{arg}(G_\mathrm{det})$ is negligible and constant at $-90^{\circ}$. The response $\mathrm{arg}(G_\mathrm{bp})$ of the bandpass filter is simply given by the additional phase delay when the filter is installed at the analog input of the phase detector. Figure \ref{fig:loopresponse}B shows the frequency-dependence of the measured gain $|G_\mathrm{det}|$ and phase delay $\mathrm{arg}(G_\mathrm{det})$. Specifically, it shows the voltage-to-voltage transfer function $G_\mathrm{det} \cdot G_\mathrm{VCO}$. As the optomechanical VCO has a bandwidth $B_\mathrm{VCO} \gg 10 \, \mathrm{kHz}$ (see Section \ref{sec:FunctionalPrinciple}), it can be regarded as a constant prefactor $G_\mathrm{VCO} \in \mathbb{R}$, which is the only free fit parameter in the loop analysis. The measurement of $G_\mathrm{PID}$ and $G_\mathrm{lp}$ is straight-forward, with the only requirement that the lowpass filter should be measured without disconnecting it from the control loop (for impedance reasons). 

Finally, the multiplication of all measured gains $|G_\mathrm{i}|$ and summing of all measured phases $\mathrm{arg}(G_\mathrm{i})$ yields the full open-loop gain $G_\mathrm{0}$. The measured closed-loop gain $|G_\mathrm{cl}|$ and phase $\mathrm{arg}(G_\mathrm{cl})$ show perfect agreement with the theoretical curves according to equation \ref{eq:G_cl} using the measured $G_\mathrm{0}$. As expected, the two closed-loop response functions converge against the open-loop responses for large frequencies, where $G_\mathrm{0}$ becomes smaller than 1. The locking bandwidth of the spring control can be defined as the frequency $f_\mathrm{0} = 370 \, \mathrm{Hz}$ where $|G_\mathrm{0}| = 1$ (zero-crossing on logarithmic scale). The lock is stable with a phase margin of $\mathrm{arg}[G_\mathrm{0}(f_\mathrm{0})] + 180^{\circ}  = 51^{\circ}$. If required it can be increased by reducing the P-gain in the PID controller, so that $f_\mathrm{0}$ is shifted further away from $f_\mathrm{\pi} = 910 \, \mathrm{Hz}$, where the open-loop phase $\mathrm{arg}(G_\mathrm{0})$ becomes $-180^{\circ}$. At this point one can read off the gain margin $-|G_\mathrm{0}(f_\mathrm{\pi})| = 8.8 \, \mathrm{dB}$, which indicates a good stability of the lock.

\textit{\textbf{Feasibility}}: The optical spring-control can be implemented in a variety of optomechanical systems in the unresolved sideband regime, where the frequency-shift of the optical spring exhibits a slope on resonance. The lock can be operated with any mechanical mode that exhibits a sufficiently large optomechanical coupling and that lies within the bandwidth of the signal processing in the lock. The advantage is that it is easy to implement and the needed elements already exist in most of the optomechanical systems.

However, some limitations of the lock should be noted that might limit the feasibility in some cases. First of all, the shape of the error signal $\Delta \omega_\mathrm{m} \left( \Delta \right)$ (see equation \ref{eq:Delta_omega}) depends only on the cavity linewidth and the mechanical frequency, which can not be tuned in general. In contrast, the shape of a PDH error signal can be easily adjusted by choice of the frequency of the sideband generation. Furthermore, the locking of the cavity length is only guaranteed, if the frequency of the mechanical oscillator is highly stable in the unperturbed case, since otherwise the frequency-shift $\Delta \omega_\mathrm{m}$ is not a defined function of the cavity detuning. Generally, this can only be achieved in cryogenic operation or with a precise temperature control. Furthermore, the functional principle of the lock implies some general drawbacks of PLLs. Firstly, the locking bandwidth is fundamentally limited by the phase detector, whose gain drops by 20 dB per decade. Secondly, the pull-in behavior of a PLL is more critical than in the case of a normal lock. In our case, we need to tune the cavity close to the desired detuning by hand, before the lock is activated. This might be difficult for a system with less passive stability. Nevertheless, we observe self-oscillations in the system when the PLL catches a sideband around $\omega_\mathrm{m}$ after a large perturbation of the MiM cavity.

\textit{\textbf{Conclusion}}: We demonstrate a novel cavity locking technique that relies on the optical spring effect to generate the error signal. 
It can be implemented in a variety of different optomechanical systems in the unresolved sideband regime. 
For characterizing the performance of the lock, we measured the closed-loop response and explain it by a detailed characterization of the open-loop response, yielding a locking bandwidth of $370 \, \mathrm{Hz}$, a phase margin of $51^{\circ}$ and a gain margin of $8.8 \, \mathrm{dB}$.

\textit{\textbf{AIP Publishing Data Sharing Policy}}: The data that support the findings of this study are available from the corresponding author upon reasonable request.

\textit{\textbf{Acknowledgement}}: We gratefully acknowledge financial support by the Cluster of Excellence "Advanced Imaging of Matter" of the German science foundation (EXC 2056). We also thank Hai Zhong and Ortwin Hellmig for technical support in earlier stages of the experiment.

\bibliography{library_submit}

\begin{thebibliography}{34}
\bibitem [{Nor()}]{Norcada}%
  \BibitemOpen
  \bibinfo {note} {Our SiN trampoline oscillator was manufactured by
  Norcada.}\BibitemShut {Stop}%
\bibitem [{G_d({\natexlab{b}})}]{G_det_2}%
  \BibitemOpen
  \bibinfo {note} {The measurement also contains the minus-sign
  ($+180^{\circ}$) of the negative feedback. It was subtracted out as it is
  explicitly not contained in $G_\mathrm{det}$.}\BibitemShut {Stop}%
\end{thebibliography}

\end{document}